\documentclass[aps,preprint]{revtex4}

\usepackage{graphicx}

\usepackage{subfigure}
\usepackage{color}
\usepackage{amsmath,amssymb}

\begin{document}
\noindent Published in: Phys. Rev. E {\bf 92}, 052106 (2015).
\title{Identifying the order of a quantum phase transition by means of  Wehrl entropy in phase-space }

\author{Octavio Casta\~nos}
\email{ocasta@nucleares.unam.mx}
\affiliation{Instituto de Ciencias Nucleares, Universidad Nacional Aut\'onoma de M\'exico, Apdo.\ Postal 70-543, 04510, DF, Mexico}
\author{Manuel Calixto}
\email{calixto@ugr.es}
\affiliation{Departamento de Matem\'atica Aplicada, Facultad de Ciencias, Universidad de Granada,
Fuentenueva s/n, 18071 Granada, Spain.}
\author{Francisco P\'erez-Bernal}
\email{francisco.perez@dfaie.uhu.es}
\affiliation{Departamento de F\'{\i}sica Aplicada, 
Facultad de Ciencias Experimentales, Universidad de Huelva, Campus del Carmen, 
Avda.\ de las Fuerzas Armadas s/n, 21071 Huelva, Spain.}
\author{Elvira Romera}
\email{eromera@ugr.es}
\affiliation{Departamento de F\'{\i}sica At\'omica, Molecular y Nuclear and Instituto Carlos I de F\'{\i}sica Te\'orica y Computacional, Universidad de Granada,
Fuentenueva s/n, 18071 Granada, Spain.}


\begin{abstract}
  We propose a method to identify the order of a Quantum Phase
  Transition by using area measures of the ground state in phase
  space. We illustrate our proposal by analyzing the well known example
  of the Quantum Cusp, and four different paradigmatic  boson
  models: Dicke, Lipkin-Meshkov-Glick, interacting boson model, and
  vibron model.
\end{abstract}

\maketitle

\section{Introduction}

The extremely relevant concept of phase transition in Thermodynamics
has been extended in later times to encompass novel situations. In
particular, two main aspects have been recently addressed: the study
of mesoscopic systems and of quantum systems at zero temperature. In
the first case, the finite system size modifies and smooths phase
transition effects. In the second case a tiny modification of
certain Hamiltonian parameter or parameters (control parameters)
induces an abrupt change in the ground state of the quantum system and
Quantum Phase Transitions (QPTs) appear as an effect of quantum
fluctuations at the critical value of the control parameter \cite{bookCarr}.  QPTs
strictly occur in infinite systems, though QPT precursors are present
in finite systems. In fact, bosonic models allow to study both
aforementioned aspects: finite-size effects and zero temperature
QPTs. Recent reviews on this subject are \cite{castenrev,cejnar2009,cejnar2010}.

QPTs occurring in finite-size systems can be characterized by the disappearance
of the gap between the ground and the first excited state energies in the
mean field or thermodynamic limit (infinite system size). The QPT is a
first order phase transition if a level crossing occurs and a continuous transition if there are no
crossings (except in the limit value) \cite{cejnar2007}. The Landau
theory holds in the models addressed in this presentation, and
within this theory the Ehrenfest classification of QPTs is valid. In
this case, the order of a QPT is assigned on the basis of
discontinuities in derivatives of the potential of the system at the
thermodynamic limit \cite{cejnar2009, cejnar2010}.

The assignment of the order of a phase transitions in finite-size
systems using a numerical treatment to compute finite differences of
the system energy functional can be a cumbersome task. In order to overcome this
problem, different approaches have been proposed.  Cejnar \textit{et
  al.} have used the study of nonhermitian
degeneracies near critical points to classify the order of
QPTs \cite{cejnar2007}. Alternative characterizations are based in the connection
between geometric Berry phases and QPTs in the case of the XY Ising
model \cite{carollo2005, zhu2006} and in the overlap between two
ground state wave functions for different values of the
control parameter (fidelity susceptibility concept)
\cite{zanardi2006,gu2010, ocasta2010}.  In addition, many efforts have been devoted to characterize
QPTs in terms of information theoretic measures of
  delocalization  (see \cite{n4, n7, n3, n42} and references therein)
and quantum information tools, e.\ g.\ using
entanglement entropy measures (see e.g.~\cite{lambert2005} for the Dicke
model and \cite{calixto2012,calixto2014} for the vibron model).

In this work we propose an alternative way to reckon the order
of a QPT by using the Wehrl entropy in the phase-space (coherent state
or Bargmann) representation of quantum states $\psi$ provided by the
Husimi function $Q_\psi$, which is defined as the squared overlap
between $\psi$ and an arbitrary coherent state. 

The Husimi function has been widely used in quantum physics, mainly in
quantum optics.  For example, the time evolution of coherent states of
light in a Kerr medium is visualized by measuring $Q_{\psi}$ by cavity
state tomography, observing quantum collapses and revivals, and
confirming the non-classical properties of the transient states
\cite{kirchmair2013}. Moreover, the zeros of this phase-space
quasi-probability distribution have been used as an indicator of the
regular or chaotic behavior in quantum maps for a variety of quantum
problems: molecular \cite{arranz2010} and atomic \cite{dando1994}
systems, the kicked top \cite{chaudhury2009}, quantum billiards
\cite{tualle1995}, or condensed matter systems \cite{weinmann1999}
(see also \cite{leboeuf1990,arranz2013} and references
therein). They have also been considered as an indicator of
metal-insulator \cite{aulbach2004} and topological-band insulator
\cite{TI} phase transitions,  as well as of QPTs in Bose
Einstein condensates \cite{ocasta2010} and in 
the Dicke \cite{real2013,romera2012}, vibron \cite{calixto2014}, and Lipkin-Meshkov-Glick (LMG) models
\cite{romera2014}.

To identify the order of a QPT we suggest to observe the singular behavior of the
Wehrl entropy, $W_\psi$, of the Husimi function, $Q_\psi$, near the
critical point as the system size increases. The Wehrl entropy,
is defined in Sec.\ \ref{Wehrlsec} as a function of the Hamiltonian control parameter(s)
and the system's size.  For harmonic oscillators, Lieb proved in  \cite{lieb1978} the Wehrl's conjecture 
\cite{wehrl1979} stating that $W_\psi$ attains its minimum (maximum area)
when $\psi$  is an ordinary (Heisemberg-Weyl) coherent state. This
proof has been recently extended by Lieb and Solovej  to SU(2)
spin-$j$ systems \cite{lieb2014}. We observe that $W_\psi$  is maximum at the
critical point of a first-order QPT, and this maximum is narrower as
the system size increases. However, for second-order QPTs, the Wehrl
entropy displays a step function behavior at the critical point, and
again the transition is sharper for larger system sizes. We shall
confirm this behavior for five models: Quantum Cusp, Dicke, LMG, a
one-dimensional realization of the interacting boson model (IBM-LMG),
and the 2D limit of the vibron model (2DVM).

We have chosen the Cusp model as a prototypical case, because this is
probably the best known catastrophe example, describing the
bifurcation of a critical point with a quartic potential. Its quantum
version \cite{gilmore1986} has been used to illustrate the effects
associated with criticality as a prior step to deal with more involved
physical situations
\cite{cejnar2008,cejnar2009,gilmore1986,emary2005}. In addition to the
Cusp model, we present results for four different realizations of
bosonic systems. The LMG model is a simple model,
  originally introduced for the description of nuclear systems as an
  exactly-solvable toy model to assess the quality of different
  approximations \cite{lipkin1965}.  This ubiquitous model still
  receives a major attention, further stimulated by its recent
  experimental realization \cite{lipkinnat1,lipkinnat2}. The study of
  the ground state quantum phase transitions for this model can be
  traced back to the seminal articles of Gilmore and Feng
  \cite{Gilmore1978,Feng1978}.  The Dicke model is a quantum-optical
model that describes the interaction of a radiation field with $N$
two-level atoms \cite{dicke1954}. This model has
  recently renewed interest
  \cite{Garraway1137,casta1,casta2,nataf}, partly because a tunable
  matter-radiation interaction is a keynote ingredient for the study
  of quantum critical effects \cite{lambert2005, emary2003,
    emary2003_2} and partly because the model phase transition has
  been observed experimentally \cite{baumann2010}.  The interacting
boson model (IBM) was introduced by Arima and
Iachello to describe the structure of low energy states of even-even
medium and heavy nuclei \cite{booknuc}. For the sake of simplicity, we
use the IBM-LMG, a simplified version of the model built with scalar
bosons \cite{vidal2006}.  Finally, the vibron model was also
proposed by Iachello to describe the rovibrational
  structure of molecules \cite{iachello1981} and the 2DVM was
  introduced \cite{iachello1996} to model molecular bending dynamics
  (e.g.\ see Ref.~\cite{Larese2013} and references therein). The 2DVM
is the simplest two-level model which still retains a non-trivial
angular momentum quantum number and it has been used as a playground
to illustrate ground state and excited state QPTs features in bosonic
models \cite{caprio2008,pbernal2008}.


We proceed to present the Hamiltonian of the five different
addressed models, defining the Wehrl entropy as functions of the moments
of the Husimi function $Q_\psi$, and the results obtained in the
first and second order critical points of the
different models considered. 
A brief introduction to the main results on
Schwinger boson realizations, coherent states, and energy surfaces used in the paper can be found in App.~\ref{appa}.




\section{Selected Models}\label{Modelssec}

We give a brief outline of the five models we use  to illustrate  the characterization
of QPT critical points by means of the Wehrl entropy.

\begin{figure}
\includegraphics[width=12cm]{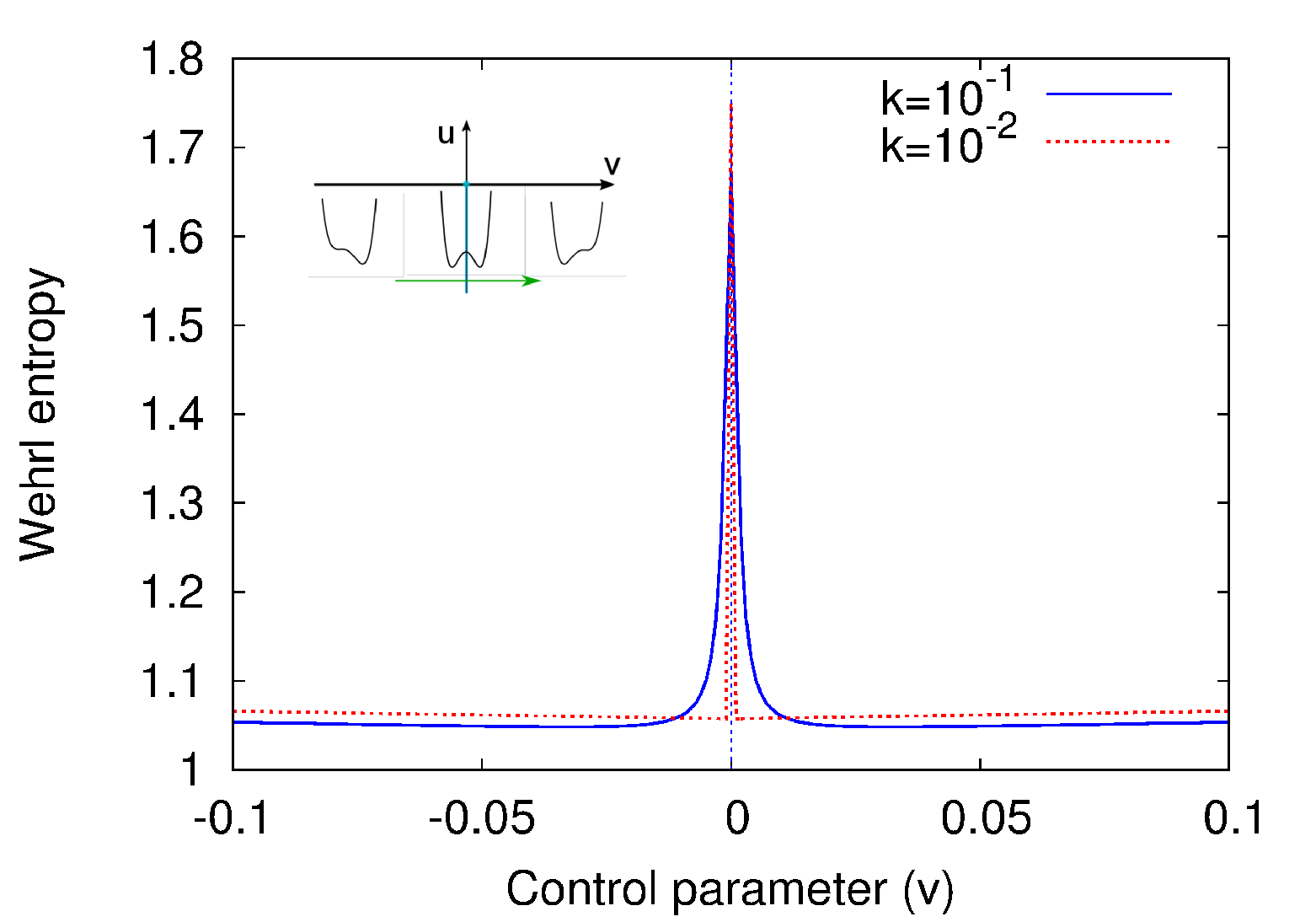}
\includegraphics[width=8.55cm,angle=-90]{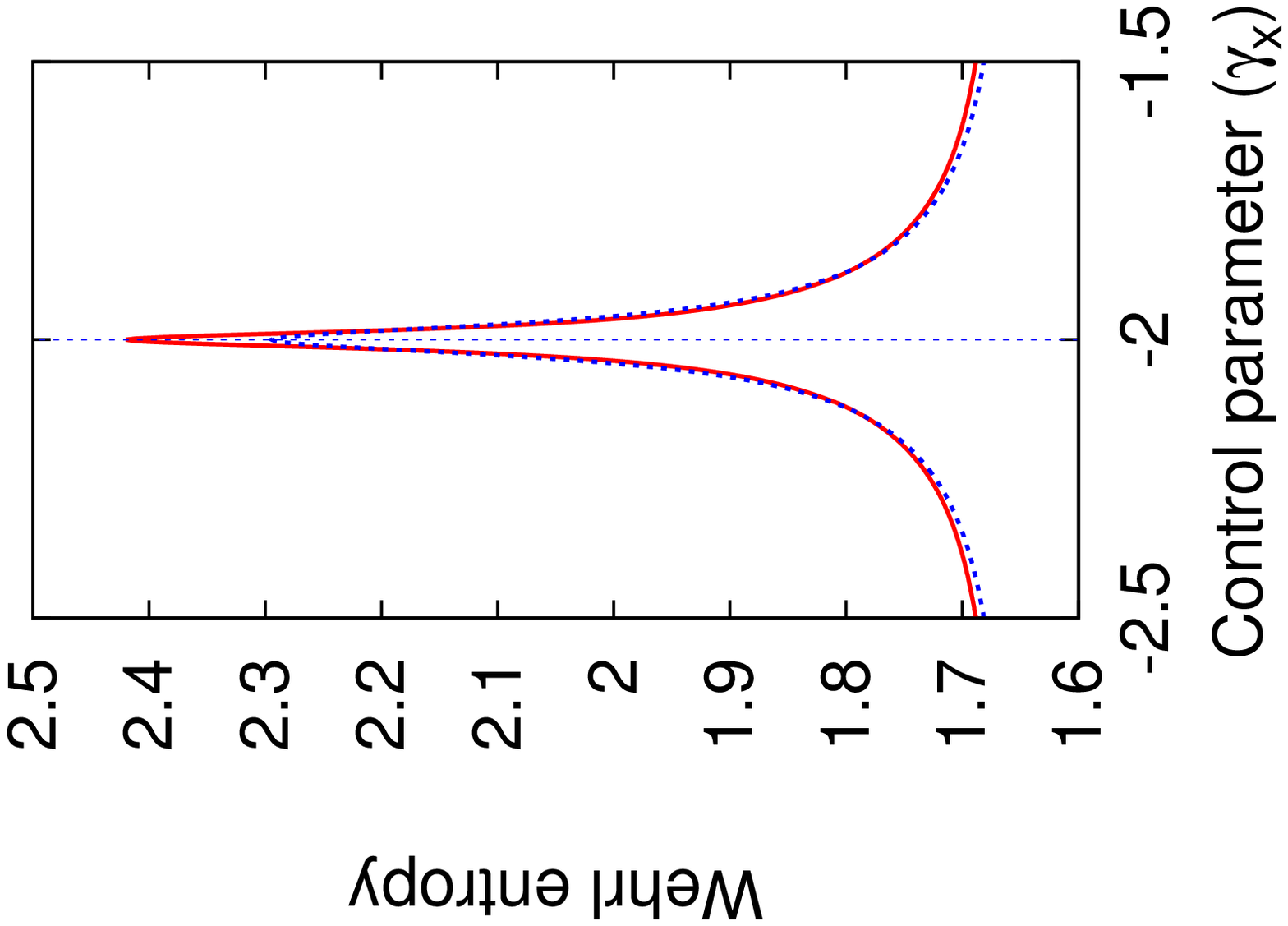}
\includegraphics[width=8.25cm,angle=-90]{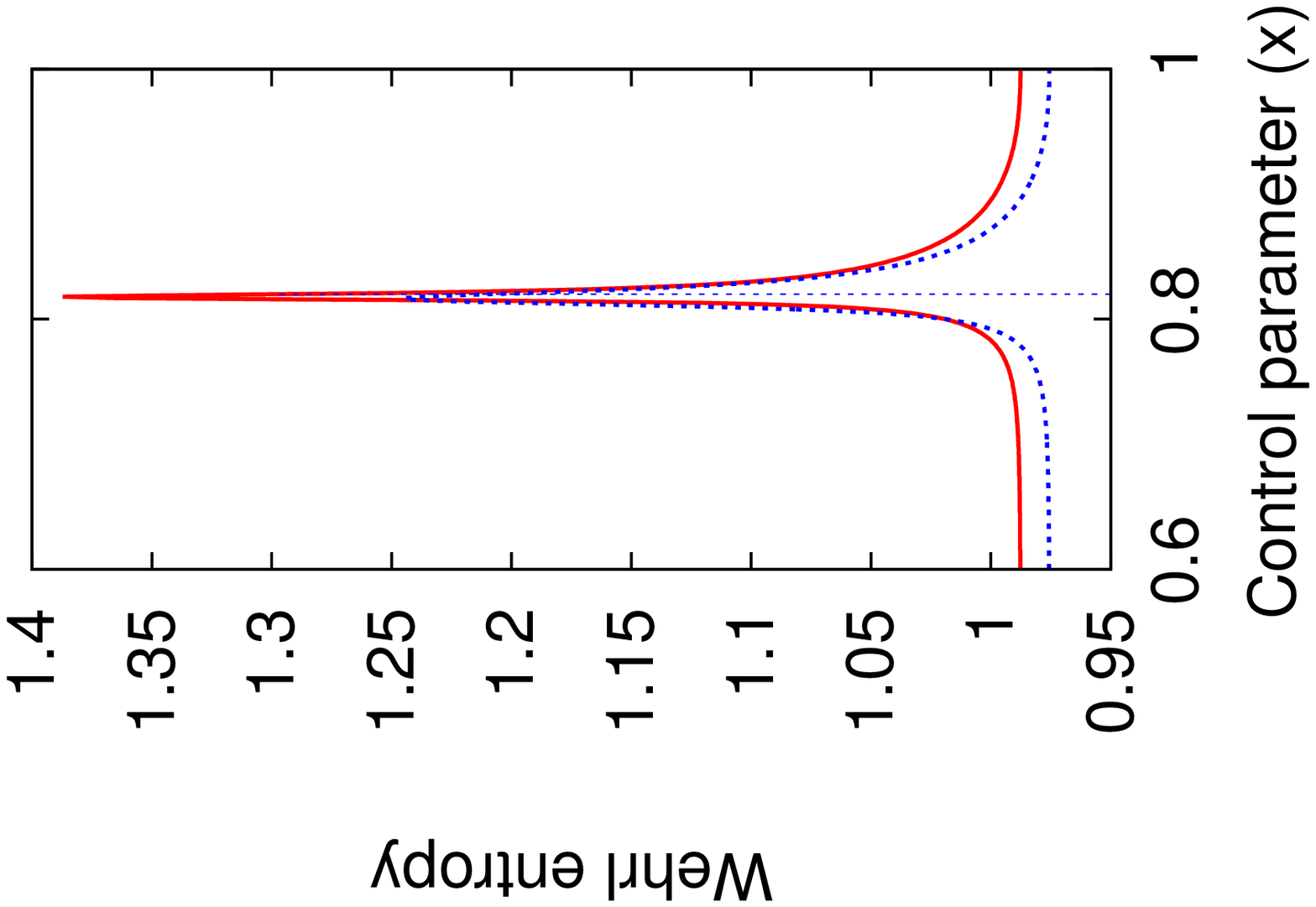}
\caption{(Color online) First order QPTs: Wehrl entropy $W_\psi$ of the Husimi
  function for the ground state. Top panel: cusp model for $k=10^{-1}$
  (blue, solid) and $10^{-2}$ (red, dashed), along the straight line
  $u=-1$ with critical point $v_c=0$. Bottom left: LMG model for
  $N=20$ (blue, dashed) and $40$ (red, solid), along the straight line
  $\gamma_x=-\gamma_y-4$ with critical point $\gamma_{x c}=-2$. Bottom
  right: IBM-LMG model for $N=80$ (red, solid) and $N=40$ (blue,
  dashed), for the straight line $y=\frac{1}{\sqrt{2}}$ with critical
  point $x_c=0.82$. Critical points are marked with vertical blue dotted lines. }\label{firstorderfig}
\end{figure}

The first model is the one dimensional quantum cusp Hamiltonian \cite{gilmore1986,cejnar2008,cejnar2009,emary2005}
\begin{equation}
 \hat H=\frac{K^2 {\hat p}^2}{2}+V_{c}(\hat x)~,
\label{cuspH}
\end{equation}
where $V_{c}(\hat x)=\frac{1}{4}{\hat x}^4+\frac{u}{2} {\hat x}^2+v
\hat x$ is the cusp potential,  with control parameters $u$ and $v$ and
a classicality constant $K=\frac{\hbar}{\sqrt{M}}$, combining $\hbar$  and the mass parameter  $M$  (see \cite{cejnar2008}). The smaller the value of $K$ the closer the system is to the classical limit. 
The  mass parameter  $M$ can be fixed to unity without loss of generality. In order to obtain
energies and eigenstates for the quantum Cusp, we have recast
Hamiltonian \eqref{cuspH} in second quantization, using harmonic oscillator
creation and annihilation operators, and diagonalized the resulting
matrix with a careful assessment of convergence. The ground state quantum phase
transitions associated with the cusp have been studied using 
Catastrophe Theory and Ehrenfest's classification \cite{cejnar2009}
and making use of entanglement singularities \cite{emary2005}. It is well known that
there is a first order quantum phase transition line when the control
parameter $v$ changes sign for negative $u$ values, and a second order
transition point for $v=0$ and $u$ moving from negative to positive
values. In this work we will consider two trajectories: (i) for $u=-1$
and $v\in[-0.2,0.2]$ with a first order critical point at $v_c = 0$,  and (ii) for $v=0$ and $u\in[-1,1]$ with a second order critical point at $u_c = 0$.

The Dicke model is an important model in quantum optics that
describes a bosonic field interacting with an ensemble of
$N$ two-level atoms with level-splitting $\omega_0$. The Hamiltonian is given by
\begin{equation}
\label{qpt01}
\hat H=\omega_0 {\hat J}_z + \omega a^{\dag} a + \frac{\lambda}{\sqrt{2 j}}
( a^{\dag} + a )( {\hat J}_+ + {\hat J}_-)~,
\end{equation}
where ${\hat  J}_z$, ${\hat  J}_{\pm}$ are angular  momentum operators
for a  pseudospin of length  $j=N/2$, and  $a$ and $a^{\dag}$  are the
bosonic   operators   of   a    single-mode   field   with   frequency
$\omega$.  There is  a second  order QPT  at a  critical value  of the
atom-field coupling strength $\lambda_c=\frac12\sqrt{\omega\omega_0}$,
with  two  phases:  the  normal phase  ($\lambda<\lambda_c$)  and  the
superradiant                phase                ($\lambda>\lambda_c$)
\cite{Hepp1973,Wang1973}.  Several
  tools  for the  identification of  its QPTs  have been  proposed: by
  means of entanglement  \cite{lambert2005}, information measures (see
  \cite{n3,n4}  and  references  therein)  and in  terms  of  fidelity
  \cite{casta3}, inverse  participation ratio,  the Wehrl  entropy and
  the    zeros    of    the     Husimi    function    and    marginals
  \cite{romera2012,real2013, hirsch2015}.

We will also deal with an interacting fermion-fermion model,
the LMG model \cite{lipkin1965}.
In the quasispin formalism, except for a constant term,
the Hamiltonian for $N$ interacting spins can be written as
\begin{equation}
\frac{\hat H}{2\omega j}=\frac{{\hat J}_z}{j}+\frac{\gamma_x}{j(2j-1)}{\hat J}^2_x+\frac{\gamma_y}{j(2j-1)}{\hat J}^2_y~,
\end{equation}
where $\gamma_x$ and $\gamma_y$ are control parameters. We would like
to point out that the total angular momentum $J^2=j(j+1)$ and the
number of particles $N=2j$ are conserved, and $\hat H$ commutes with
the parity operator for fixed $j$. Ground state quantum phase transitions
  for this model have been characterized using the continuous unitary
  transformation technique \cite{Dusuel2004}, investigating
  singularities in the complex plane (exceptional points)
  \cite{Heiss2005}, and from a semiclassical perspective
  \cite{Leyvraz2005}. A complete classification of the critical points
  has been accomplished using the catastrophe formalism
  \cite{castanos05,ocasta1}.  We will study the first and second
order QPTs given by the trajectories $\gamma_x=-\gamma_y-4$ and
$\gamma_x=-\gamma_y+2$ in the phase diagram
\cite{ocasta1}. A characterization of QPTs in the
  LMG model has recently been performed in therms of R\'enyi-Wehrl
  entropies, zeros of the Husimi function and fidelity and fidelity
  suspceptibility concepts \cite{romera2014}.

In the case of the characterization of the phase
  diagram associated with the IBM it is important to emphasize the
  pioneer works on shape phase transitions on nuclei \cite{Feng1981},
  that anticipated the detailed construction of the phase diagram of
  the interacting boson model using either catastrophe theory
  \cite{Feng1981,Lopez1996}, the Landau theory of phase transitions
  \cite{Iachello1998,Jolie2002}, or excited levels repulsion and
  crossing \cite{Arias2003}. In the present work we use the IBM-LMG, a
  simplified one dimensional model, which shows first and second order
  QPTs, having the same energy surface as the Q-consistent interacting
  boson model Hamiltonian \cite{vidal2006}.  In this case the
Hamiltonian is
\begin{equation}
\hat H=x {\hat n}_t-\frac{1-x}{N}{\hat Q}^y {\hat Q}^y~,
\end{equation}
with ${\hat n}_t=t^{\dag} t$ and ${\hat Q}^y=s^{\dag} t + t^{\dag} s +
y\, t^{\dag} t$ are expressed in terms of two species of scalar bosons
$s$ and $t$, and the Hamiltonian has two control parameters $x$ and $y$. The total
number of bosons $N= {\hat n}_s+ {\hat n}_t$ is a conserved quantity.
For $y=0$ there is an isolated point of second order phase transition
as a function of $x$ with a critical value $x_c = 0.8$. For $y>0$ the
phase transition is of first order and, to illustrate this case, we
have chosen the value $y = 1/\sqrt{2}$, with a critical control
parameter $x_c = 0.82$.

Finally, the 2DVM is a model which describes a system containing a
dipole degree of freedom constrained to planar motion.  Elementary
excitations are (creation and annihilation) 2D Cartesian $\tau$-bosons
and a scalar $\sigma$-boson. The second order ground state quantum phase transition in this model has been studied in Ref.~\cite{pbernal2008} 
using the essential Hamiltonian
\begin{equation}
 \hat{H}=(1-\xi)\hat{n}+\xi\frac{N(N+1)-\hat{W}^2}{N-1},\label{hamiltonian}
\end{equation}
where the (constant) quantum number $N$ labels the totally
symmetric 
representation $[N]$ of U(3),
$\hat{n}=\tau_+^\dag\tau_++\tau_-^\dag\tau_-$ is the number operator
of vector bosons, and
$\hat{W}^2=(\hat{D}_+\hat{D}_-+\hat{D}_-\hat{D}_+)/2+\hat{l}^2$. The
operators $\hat{D}_+=\sqrt{2}(\tau^\dag_+\sigma-\sigma^\dag\tau_-)$
and $\hat{D}_-=\sqrt{2}(-\tau^\dag_-\sigma+\sigma^\dag\tau_+)$ are
dipole operators, and $\hat{l}=\tau_+^\dag\tau_+-\tau_-^\dag\tau_-$ is
the angular momentum operator. This model has a
  single control parameter $0\le\xi\le 1$ and the second order QPT
  takes place at a critical value $\xi_c=0.2$ \cite{pbernal2008}. Several procedures have been used
  to identify the ground state QPT in this model: entanglement entropies
  \cite{calixto2012}, R\'enyi entropies \cite{n7}, the Wehrl entropy,
  and the inverse participation ratio of the Husimi function
  \cite{calixto2012b}.

\begin{figure}
\includegraphics[width=9cm]{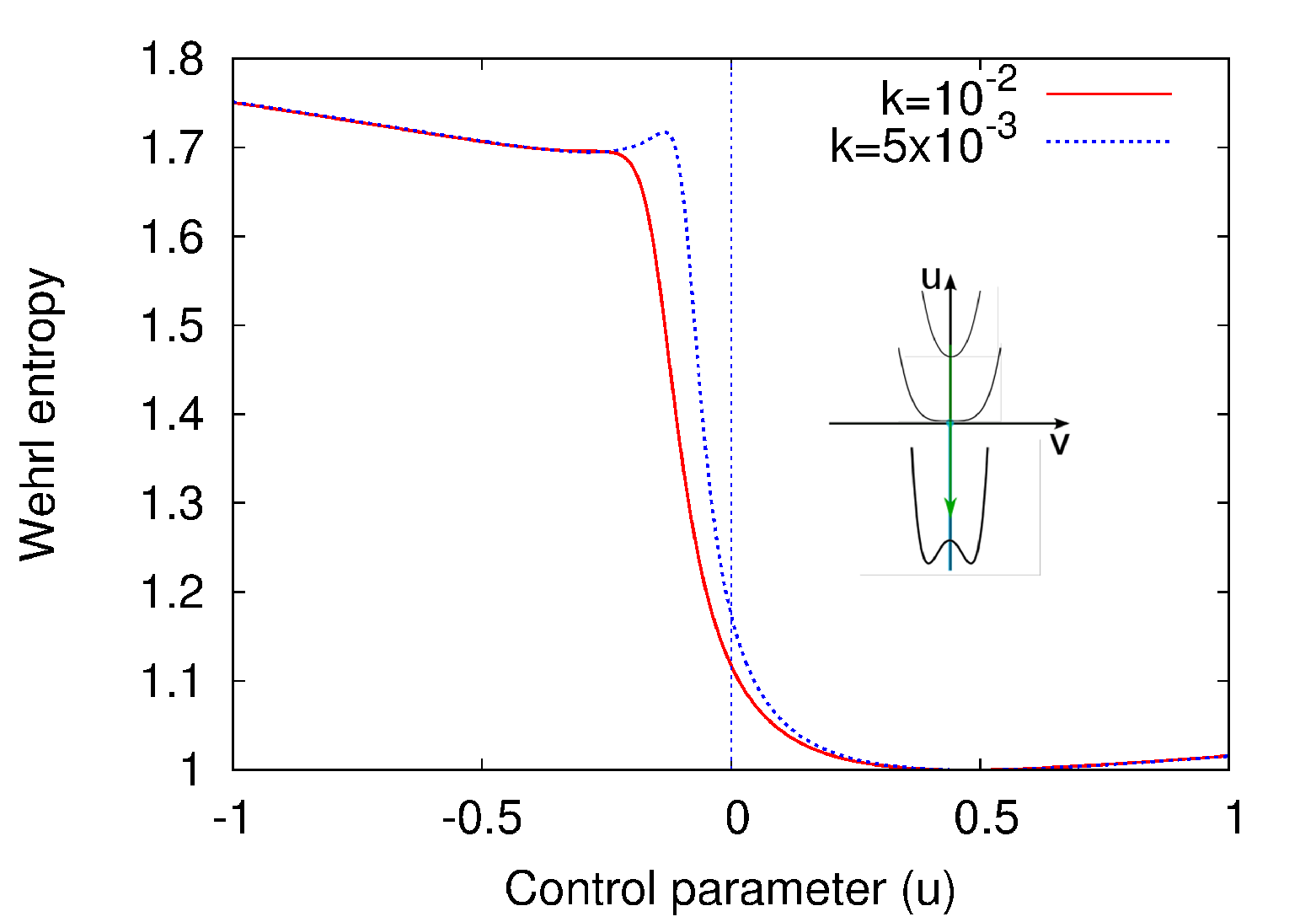}\\
\includegraphics[width=6.6cm,angle=-90]{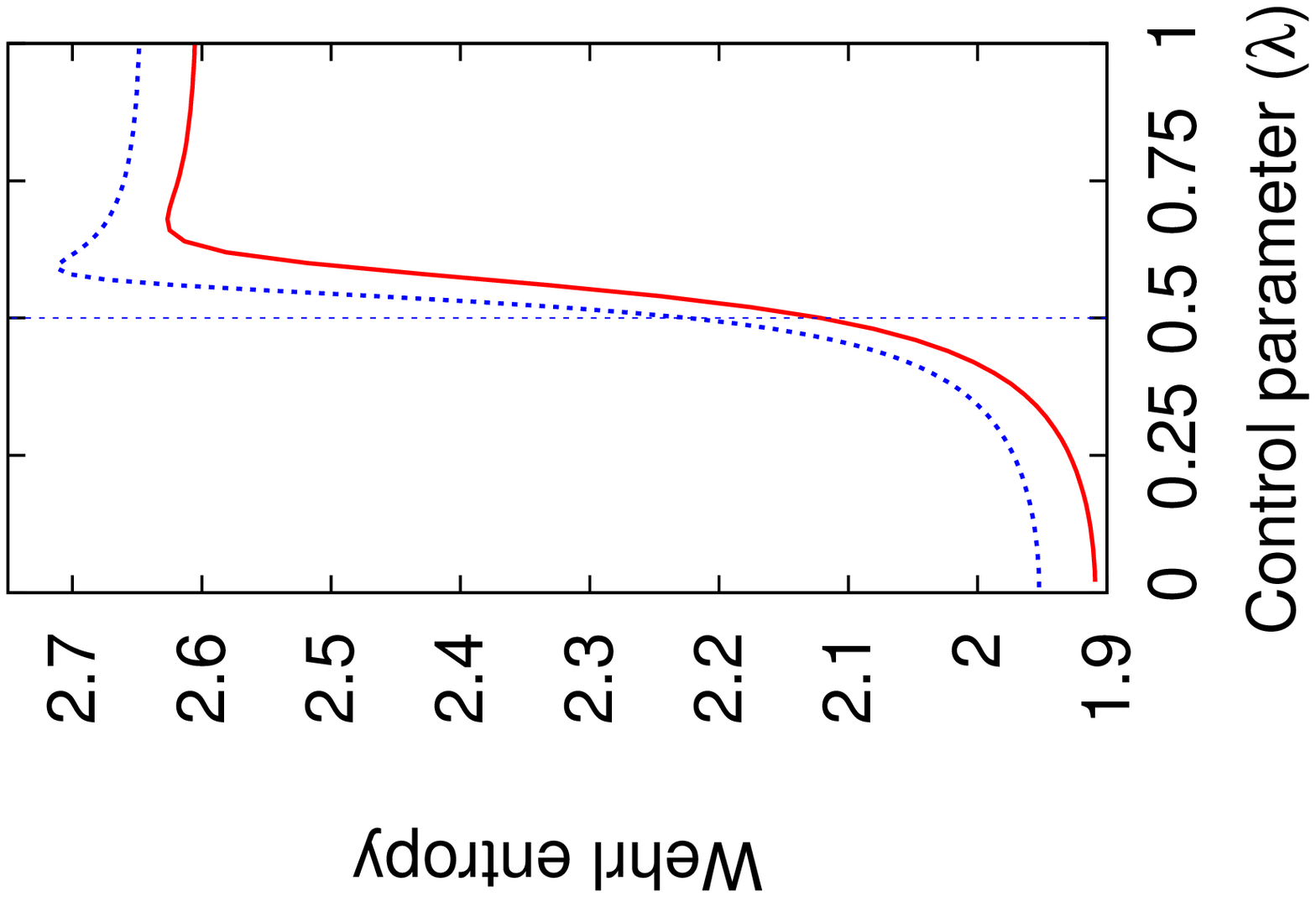}
\includegraphics[width=6.6cm,angle=-90]{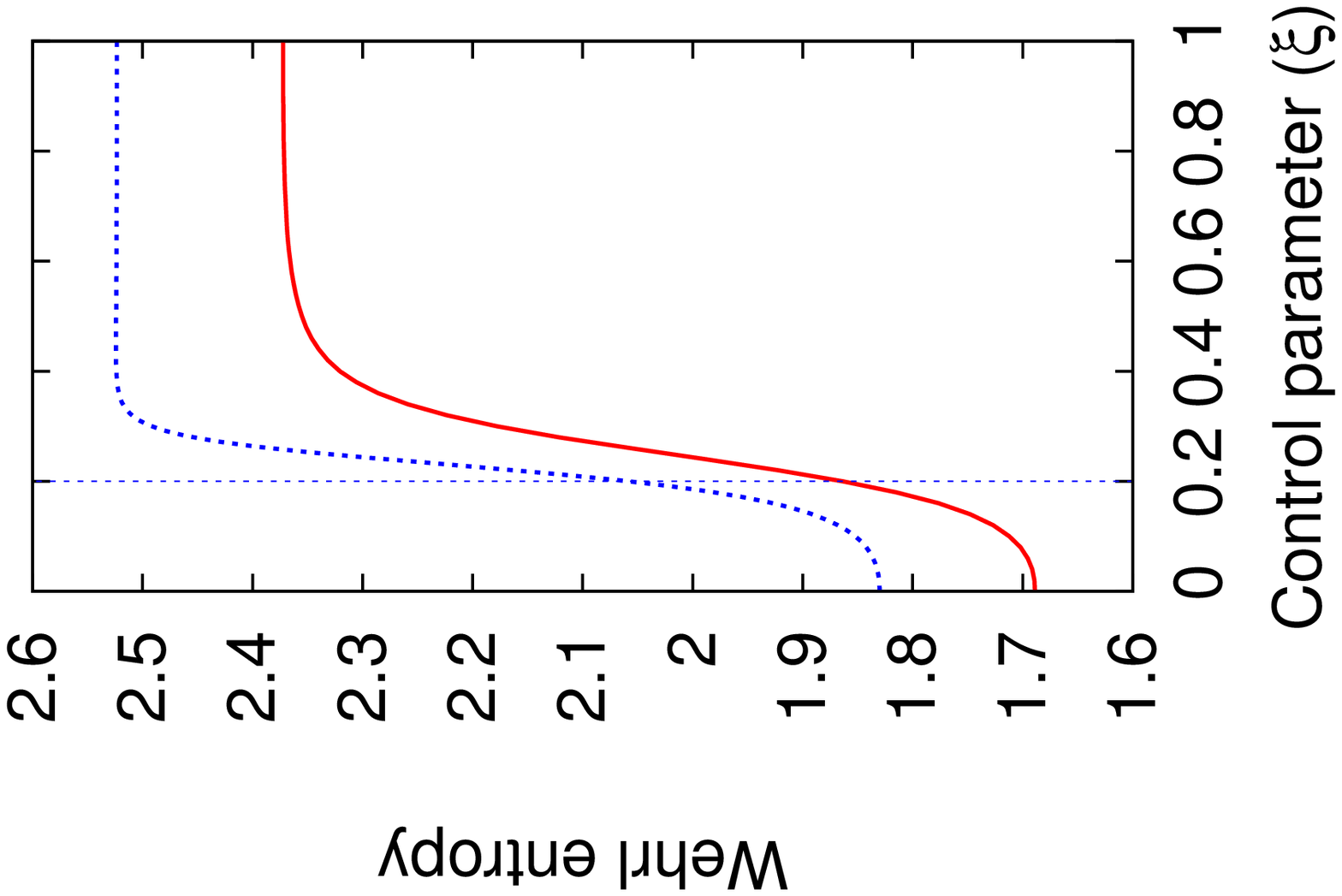}\\
\includegraphics[width=6.84cm,angle=-90]{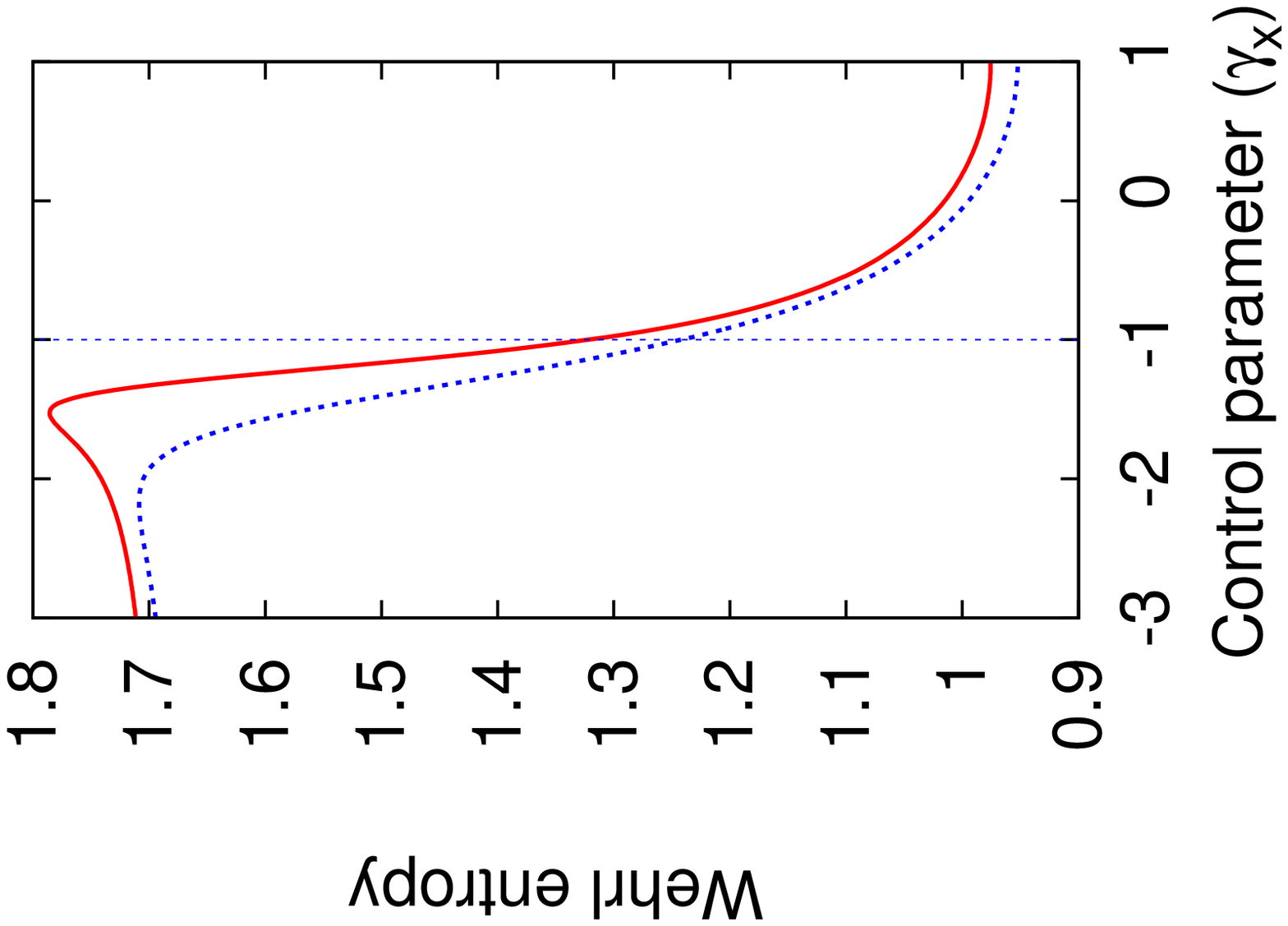}
\includegraphics[width=6.6cm,angle=-90]{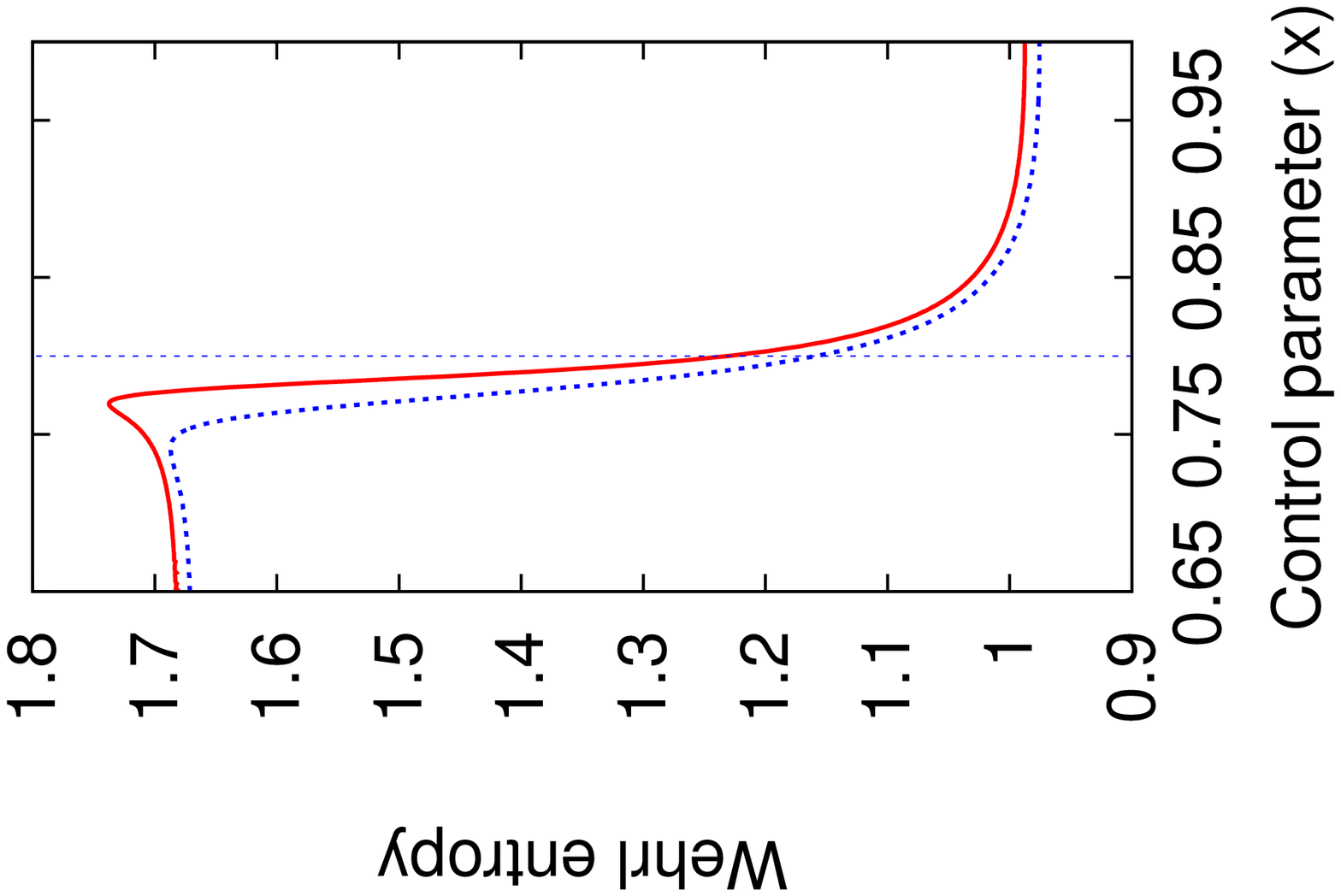}
\caption{(Color online) Second order QPTs: Wehrl entropy $W_\psi$ of the Husimi function for the ground state. Top panel: cusp model for $K=10^{-2}$ (red, solid) and $10^{-3}$
(blue, dashed), along the
straight line $u=0$ with critical point $v_c=0$. Mid left panel: Dicke model
for $N=10$ (red, solid) and $20$ (blue, dashed) with critical point $\lambda_c=0.5$; mid right panel:
2DVM results for $N=8$ (red, solid) and $16$ (blue, dashed) with critical point $\xi_c=0.2$. Bottom left panel:
LMG model for $N=20$ (blue, dashed) and $40$ (red, solid), along the straight line
$\gamma_x=-\gamma_y+2$ with critical point $\gamma_{x c}=-1$. Bottom right panel:
IBM-LMG model for $N=80$ (red, solid) and $N=40$ (blue, dashed), for the straight line $y=0$ with critical point $x_c=0.8$.}
\label{secondorderfig}
\end{figure}

\section{Wehrl's entropy and ground state QPTs}\label{Wehrlsec}

We have numerically diagonalized the Hamiltonians of the five models
for two different values of the system size $N$ in an interval of control parameters
containing a critical point (either first- or second-order). Given the expansion $|\psi\rangle=\sum_n c_n|n\rangle$ of the ground state in a basis $\{|n\rangle, n\in I\}$
($I$ denotes a set of quantum indices) with coefficients $c_n$ depending on the control parameters and the system's size $N$, and given
the expansions  of coherent states  $|\zeta\rangle$ in the corresponding basis (see App.~\ref{appa}),
we can compute the Husimi function $Q_\psi(\zeta)=|\langle \zeta|\psi\rangle|^2$ and the Wehrl entropy
\begin{equation}
  W_\psi=-\int Q_\psi(\zeta)\ln[Q_\psi(\zeta)] d\mu(\zeta),
 \end{equation}
 where we are generically denoting by $d\mu(\zeta)$ the measure in
 each phase space with points labeled by $\zeta$. Note that $W_\psi$
 is a function of the control parameters and the system size $N$. We
 discuss typical (minimum) values of $W_\psi$ for each model, which
 are attained when the ground state $\psi$ is coherent itself, and
 Wehrl entropy values of parity-adapted (Schr\"odinger cat) coherent states \cite{cat1,cat2},
 which usually appear in second-order QPTs \cite{calixto2012,calixto2014,real2013,romera2014,casta1,casta2}.

 \noindent {\it Cusp}: in the top panel of Figs.\ \ref{firstorderfig}
 and \ref{secondorderfig} we plot  $W_\psi$ as a function of
 the control parameters $u$ and $v$ for two trajectories and two
 values of the classicality constant $K$. The first order case is for
 trajectory $u=-1$, depicted in Fig.\ \ref{firstorderfig}, with a
 critical control parameter $v_c=0$. In this case it is immediately
 apparent a sudden growth of Wehrl entropy of the ground state at the
 critical point $v_c=0$.  The entropy growth is sharper as $K$
 decreases. The ground state is approximately a coherent state for
 $v\not=0$ and a cat-like state for $v=0$. Indeed, as conjectured by Wehrl
\cite{wehrl1979} and proved by Lieb \cite{lieb1978}, any Glauber
coherent state $|\psi\rangle=|\beta\rangle$ has a minimum Wehrl
entropy of $W_\psi=1$. It has also been shown
\cite{romera2012,calixto2012,real2013,romera2014} that parity adapted
coherent (Schr\"odinger cat) states,
$|\psi\rangle\propto|\beta\rangle+|-\beta\rangle$, increase the
minimum entropy by approximately $\ln(2)$ (for negligible overlap
$\langle -\beta|\beta\rangle$).  With this information, we infer that
the ground state $|\psi\rangle$ is approximately a coherent state in
the phase $u>0$ and a cat-like state in the phase $u<0$. 

The
second order QPT case is shown in Fig.\ \ref{secondorderfig}, with
$v=0$ and critical control parameter $u_c=0$.  For the
second trajectory, if we move from positive to negative values of $u$,
we find in the top panel of Fig.\ \ref{secondorderfig} a sudden growth
of $W_\psi$ in the vicinity of the critical point $u_c=0$ jumping from $W_\psi(u>0)\simeq
1$ to $W_\psi(u<0)\simeq 1+\ln(2)$. The entropy growth is sharper as
$K$ decreases (classical limit).  

Therefore, we would like to emphasize the utterly different entropic
behavior of first- and second-order QPTs. In both cases we also plot an
inset with the parameter trajectory and the evolution of the potential
along it. We proceed to show that this Wehrl entropy behavior is
shared by the rest of the considered models too, allowing a clear
distinction between first and second order QPTs.


\noindent {\it LMG}: the LMG model has first and second order
transitions depicted in the bottom left panels of Figs.\
\ref{firstorderfig} and \ref{secondorderfig}, respectively. We plot
$W_\psi$ as a function of the control parameters $\gamma_x$ and
$\gamma_y$ for the trajectories: $\gamma_y=-\gamma_x-4$ (1$^{st}$
order QPT at $\gamma_{xc}=-2$, bottom left panel Fig.\
\ref{firstorderfig}) and $\gamma_y=-\gamma_x+2$ (2$^{nd}$ order QPT at
$\gamma_{xc}=-1$, bottom left panel Fig.\ \ref{secondorderfig}), for
two values of the total number of particles $N$.  We observe an
entropic behavior completely similar to the Cusp model. The difference
only lies on the particular entropy values. In fact, according to
Lieb's conjecture \cite{lieb1978,lieb2014}), spin-$j$ coherent states
have a minimum Wehrl entropy of $W_\psi=\frac{2j}{2j+1}$, which tends
to $W_\psi=1$ in the thermodynamic limit $j\to\infty$. Cat-like states again
increase the minimum entropy by approximately $\ln(2)$.  The IBM-LMG
model exhibits a similar behavior to the LMG model, as can be
appreciated in the bottom right panel of Figs.\ \ref{firstorderfig}
and \ref{secondorderfig}.

\noindent {\it Dicke}:  the Dicke model exhibits a 2nd-order QPT at the critical value of the control parameter $\lambda_c=0.5$, when going from the normal ($\lambda<\lambda_c$) to the superradiant ($\lambda>\lambda_c$) phase.
$W_\psi$ captures this transition, as it can be seen in the mid left panel of Fig.\  \ref{secondorderfig}, showing an entropy increase from
$W_\psi\simeq 1+\frac{N}{N+1}$ to $W_\psi\simeq
1+\frac{N}{N+1}+\ln(2)$, with $N=2j$ the number of atoms. As expected,
the entropic growth at $\lambda_c$ is sharper for higher $N$.


\noindent {\it Vibron}: the vibron model undergoes a 2nd-order (shape) QPT
at $\xi_c=0.2$, the critical point that marks a change between linear
($\xi<\xi_c$) and bent ($\xi>\xi_c$) phases \cite{pbernal2008}.  In
the mid right panel of Fig.\ \ref{secondorderfig} we plot the Wehrl
entropy as a function of $\xi$ for two values of the system's size $N$
(total number of bosons). As in the previous models, the 2nd-order QPT
is characterized by a ``step function'' behavior of $W_\psi$
near the critical point. In this case, we have conjectured
\cite{calixto2012} that minimum entropy
$W_\psi=\frac{N(3+2N)}{(N+1)(N+2)}$ is attained for U(3) coherent
states.  In the bent phase, the ground state $|\psi\rangle$ is a cat
\cite{calixto2012,calixto2012b,calixto2014} and therefore
$W_\psi\simeq \frac{N(3+2N)}{(N+1)(N+2)}+\ln(2)$.

\section{Concluding remarks}\label{Concsec}

In summary, we have numerically diagonalized the Hamiltonians of five
models for several system's sizes $N$ in a given interval of control
parameters that contains a critical point (either of first or second
order). Given the expansion $|\psi\rangle=\sum_n c_n|n\rangle$ of the
ground state in a basis $\{|n\rangle, n\in I\}$ ($I$ denotes a set of
quantum indices) with coefficients $c_n$ depending on the control
parameters and the system's size $N$, and given the expansions
 of coherent states in the
corresponding basis, we can compute the Husimi function $Q_\psi$ and
the Wehrl entropy $W_\psi$. In Figs.\ \ref{firstorderfig} and
\ref{secondorderfig} we plot $W_\psi$ as a function of a control
parameter for different values of $N$.

From the obtained results it is
clear that the  Wehrl entropy behavior at the vicinities of
the critical point is an efficient numerical way of distinguishing
first order and continuous QPTs.

It is worth to emphasize that the present approach could imply an
extra computational cost if compared to the search of nonanaliticities
in the ground state energy functional. The present method makes use of
the ground state wave functions for different values of the control
parameter and it also requires the calculation of the overlap of the
basis states with the coherent states.  Though the need of ground
state wavefunctions instead of ground state energies is
computationally more exigent, the finer sensitivity of the present
method largely offsets the extra computational cost. The second step,
the overlap with coherent states, needs to be done only once with
available analytic expressions (see App.~\ref{appa}), therefore it
does not constitute a significant computational burden. The proposed
approach permits a clear determination of the character of a critical
point using relatively small basis sets. On the contrary, even for
large system sizes, the numerical determination with finite
differences of the critical points character could remain ambiguous.

A similar sensitivity and computational cost could be attained with
the fidelity susceptibility approach, that provides a clear
determination of the critical point location, but with no information
of the transition order and with the additional hindrance of varying
the control parameter in two different scales. Something similar
happens with entanglement entropy measures, that are suitable to be
applied to bipartite or multipartite systems, the critical point is
clearly located but no precise information about the transition order
is obtained.

\begin{acknowledgments}
We thank J.\ E.\ Garc\'{\i}a Ramos
  for useful discussion. Work in
University of Huelva was funded trough MINECO grants
FIS2011-28738-C02-02 and FIS2014-53448-C2-2-P and by Spanish Consolider-Ingenio 2010 (CPANCSD2007-00042). Work in University of Granada was supported by the  Spanish Projects:  
MINECO FIS2014-59386-P, and the Junta de Andaluc\'{\i}a projects P12.FQM.1861 and FQM-381.
\end{acknowledgments}

\appendix

\section{\label{appa}Schwinger boson realizations, coherent states  and energy surfaces}

\paragraph{Single mode} radiation fields are described by  harmonic oscillator creation $a^\dag$ and annihilation 
$a$ operators in Fock space $\{|n\rangle=\frac{(a^\dag)^n}{\sqrt{n!}}|0\rangle\}$, 
and the corresponding normalized coherent state (CS) is given by:
\begin{equation}
 |\alpha\rangle=e^{-|\alpha|^2/2}e^{\alpha a^\dag}|0\rangle=e^{-|\alpha|^2/2}\sum_{n=0}^\infty 
 \frac{\alpha^n}{n!}|n\rangle,\label{cohHW}
\end{equation}
where $\alpha=x+i p\in\mathbb C$ is given in terms of the quadratures $x, p$ of the field. 
The phase space (Bargmann) representation of a given normalized state 
$|\psi\rangle=\sum_{n=0}^\infty c_n|n\rangle$ of the 
(single mode) radiation field is given by the Husimi function $Q_\psi(\alpha)=|\langle \alpha|\psi\rangle|^2$, 
which is normalized according to $\int_{\mathbb{R}^2} Q_\psi(\alpha)d\mu(\alpha)=1$, with measure $d\mu(\alpha)=\frac{1}{\pi}d^2\alpha=\frac{1}{\pi}dxdp$.

\paragraph{Two-mode} ($a_1, a_2$) boson condensates with $N=2j$ particles 
are described in terms of SU(2) operators, whose Schwinger realization 
is
\begin{equation}
 J_+=a_2^\dag a_1, \, J_-=a_1^\dag a_2, \, J_z=\frac{1}{2}(a_2^\dag a_2-a_1^\dag a_1).\label{swsu2}
\end{equation}
In the case of the Dicke model, $J_\pm, J_z$ represent collective operators for an ensemble of $N$ two-level atoms. 
Spin-$j$ coherent states (CSs) are written in terms of the Fock basis states $|n_1=j-m; n_2=j+m\rangle\equiv |j,m\rangle$ 
(with $n_1$ and $n_2$ the occupancy number of levels $1$ and $2$) as:
\begin{eqnarray}
&& |\zeta\rangle=\frac{1}{\sqrt{(2j)!}}
\frac{(a_2^{\dag}+ \zeta a_1^{\dag})^{2j}}{(1+|\zeta|^2)^j}|0\rangle\nonumber\\
&=&(1+|\zeta|^2)^{-j}\sum_{m=-j}^j{\binom{2j}{j+m}}^{1/2}\zeta^{j+m}|j,m\rangle,
\label{cohsu2}
\end{eqnarray}
where $\zeta=\tan(\theta/2)e^{-i\phi}$ is given in terms of the polar $\theta$ and azimuthal $\phi$ 
angles on the Riemann sphere. The phase-space representation of a normalized state 
$|\psi\rangle=\sum_{m=-j}^j c_m|j,m\rangle$ is now  $Q_\psi(\zeta)=|\langle \zeta|\psi\rangle|^2$, which is 
normalized according to $\int_{\mathbb S^2} {Q_\psi}(\zeta) d\mu(\zeta)=1$, 
with integration measure (solid angle) 
$d\mu(\zeta)=\frac{2j+1}{4\pi}\sin\theta d\theta d\phi$.

The IBM-LMG model, based on a scalar ($s$) and a pseudo-scalar ($t$)
boson creation and annihilation operators has been written in terms of
SU(2) operators \eqref{swsu2}, with $s=a_1$ and $t=a_2$.

\paragraph{Three-mode} ($a_0, a_1, a_2$) models (like the 2DVM) with $N$ particles 
are described in terms of U(3) operators, whose Schwinger realization 
is $J_{jk}=a^\dag_j a_k, j,k=0,1,2$. U(3) coherent states, in the symmetric representation, 
are written in terms of the Fock basis states $|n_0=N-n; n_1=(n+l)/2; n_2=(n-l)/2\rangle\equiv |N,n,l\rangle$ 
[with $n_j$ the occupancy number of level $j=0,1,2$ and $n=0,\dots,N$ (the bending quantum number), $l=n-2m$ (the 2D angular momentum), $m=0,\dots, n$] as:
\begin{eqnarray}
|\zeta_1,\zeta_2\rangle&=&\frac{1}{\sqrt{N!}}
\frac{(a_0^{\dag}+ \zeta_1 a_1^{\dag}+ \zeta_2 a_2^{\dag})^{N}}{(1+|\zeta_1|^2+|\zeta_2|^2)^{N/2}}|0\rangle,
\nonumber\\
&=&\sum_{n=0}^N\sum_{m=0}^n\frac{\{N!/[(N-n)!(n-m)!m!]\}^{1/2}}{ (1+|\zeta_1|^2+|\zeta_2|^2)^{N/2}} 
\nonumber\\ && \times\zeta_1^{n-m}\zeta_2^m|N,n,l=n-2m\rangle,
\label{cohsu3}
\end{eqnarray}
with $\zeta_1,\zeta_2\in \mathbb C$. 
The phase-space representation of a normalized state 
$ |\psi\rangle=\sum_{n=0}^{N}\sum_{m=0}^{n} c_{nm}|N,n,l=n-2m\rangle$ is now  
$Q_\psi(\zeta_1,\zeta_2)=|\langle \zeta_1,\zeta_2|\psi\rangle|^2$, which is 
normalized according to $ \int_{\mathbb R^4}  Q_{\psi}(\zeta_1,\zeta_2) d\mu(\zeta_1,\zeta_2)=1$, where 
\begin{equation}
 d\mu(\zeta_1,\zeta_2)=\frac{(N+1)(N+2)}{\pi^2}\frac{d^2 \zeta_1d^2\zeta_2}{(1+|\zeta_1|^2+|\zeta_2|^2)^3}\nonumber
\end{equation}
is the integration measure on the complex projective (quotient) space $\mathbb CP^2=$U(3)/U(2)$\times$U(1) and 
$d^2\zeta_{1,2}\equiv d\mathrm{Re}(\zeta_{1,2})d\mathrm{Im}(\zeta_{1,2})$ the usual Lebesgue measure 
on $\mathbb R^2$. 

The connection with our U(3) construction to the 2DVM is $a_0=\sigma$ and $a_{1,2}$ are the so called circular bosons: $\tau_\pm=\mp(\tau_x\mp
i\tau_y)/\sqrt{2}$, respectively.

In order to make the article as self-contained as possible, let us also briefly recall the classical Hamiltonians or energy surfaces (the Hamiltonian operator expectation value in a coherent state) and 
their critical points for the selected models. The cusp model has already been discussed in section \ref{Modelssec}.

For the Dicke model, using harmonic oscillator CSs \eqref{cohHW} for the field and 
spin-$j$ CSs \eqref{cohsu2} for the atoms, the energy surface turns out to be:
\begin{equation}
  \langle \alpha, \zeta|\hat H|\alpha,\zeta\rangle
  =\omega|\alpha|^2+j\omega_0\frac{|\zeta|^2-1}{|\zeta|^2+1}+{\lambda}{\sqrt{2j}}\frac{4\Re(\alpha)\Re(\zeta)}{|\zeta|^2+1}~.
\end{equation}

Minimizing with respect to $\alpha$ and $\zeta$ gives the equilibrium points $\alpha_e=0=\zeta_e$ if $\lambda<\lambda_c$ (normal phase) and 
\begin{equation}
\alpha_e=
-\sqrt{2j}\sqrt{\frac{\omega_0}{\omega}}\frac{\lambda}{\lambda_c}\sqrt{1-\frac{\lambda_c^4}{\lambda^4}}, \; 
\zeta_e=
\sqrt{\frac{\lambda^2-\lambda_c^2}{\lambda^2+\lambda_c^2}},
\label{critpoints}
\end{equation}
\noindent if $\lambda\geq\lambda_c$ (superradiant phase). 
For the LMG model, the energy surface written in terms of  $\zeta=\tan(\theta/2)e^{-i\phi}$ is:
\begin{equation}
\frac{\langle \zeta|\hat H|\zeta\rangle}{2 \omega j }=- \, \cos\theta+ \sin^2\theta(\frac{\gamma_x}{2} \cos^2\phi +
\frac{ \gamma_y}{2}\sin^2\phi).\label{surfaceenergy}
\end{equation}
The minimization process results in three phases for this system: 1) region $\gamma_{x} < -1$ with $\gamma_{x} < \gamma_{y}$, 2) region  $\gamma_{y} < -1$ with  
$\gamma_{y} < \gamma_{x}$ and 3) region $\gamma_{y} > -1$ and $\gamma_x>-1$; for more information, like bifurcation sets associated with the
absolute minimum of the energy surface, we address the reader to  Ref.~\cite{ocasta1}.

The analysis of the IBM-LMG case performed in \cite{vidal2006} shows how for a two-mode coherent state $|\beta\rangle$ [the large $N$ limit of $|\zeta\rangle$ in \eqref{cohsu2},  with $\zeta=\beta\in \mathbb{R}$], the resulting energy surface in the thermodynamic limit is
\begin{align}
\frac{\langle \zeta|\hat H|\zeta\rangle}{N}=&\frac{\beta^2}{(1+\beta^2)^2}\label{surfaceenergyIBM}\\
\times&\left\{5x - 4 + 4\beta y(x-1)+\beta^2\left[x + y^2(x-1)\right]\right\}~,\nonumber
\end{align}
that coincides with that of the Q-consistent IBM Hamiltonian
\cite{vidal2006}. If the control parameter $y = 0$ there is an
isolated second order phase transition point as a function of the
control parameter $x$ with a critical value $x_c = 0.8$. If $y > 0$
and constant the phase transition is of first order and minima
coexistence occurs for the critical value $x_c = (4+y^2)/(5+y^2)$. In
particular, the results shown for a first order phase transition in
the bottom right panel of Fig.~\ref{firstorderfig}, with $y =
1/\sqrt{2}$, are equivalent to the results obtained in the IBM model
in the case of a transition from a U(5) (spherical) to a SU(3) (axially symmetric) configuration in the Casten triangle \cite{Iachello1998}.

Finally, for the 2DVM \cite{pbernal2008}, due to the underlying (rotational) symmetries, one can restrict himself to particular U(3) CSs \eqref{cohsu3} with $\zeta_1=r/\sqrt{2}=-\zeta_2$, so that the energy surface turns out to be simply
\begin{eqnarray}
\frac{\langle\zeta_1,\zeta_2|\hat{H}|\zeta_1,\zeta_2\rangle}{N}=
(1-\xi)\frac{r^2}{1+r^2}+\xi
\left(\frac{1-r^2}{1+r^2}\right)^2.\label{energyns}
\end{eqnarray}
The minimization process results in two phase-shapes: 1) linear phase ($\xi\leq \xi_c=1/5$), with 
 `equilibrium radius' $r_e=0$ and 2) bent phase  ($\xi> \xi_c$), with $r_e(\xi)=\sqrt{({5\xi-1})/({3\xi+1})}$.

\end{document}